\newcounter{myctr}
\def\myitem{\refstepcounter{myctr}\bibfont\noindent\ifnum\themyctr>9\else\phantom{0}\fi\hangindent17pt\themyctr.\enskip}

\documentclass{ws-ijqi}

\begin{document}

\markboth{Suranjana Ghosh and Irene Marzoli}
{Super revivals and sub-Planck scale structures of a slightly relativistic particle in a box}

\catchline{}{}{}{}{}

\title{SUPER REVIVALS AND SUB-PLANCK SCALE STRUCTURES \\
       OF A SLIGHTLY RELATIVISTIC
       PARTICLE IN A BOX 
      }

\author{SURANJANA GHOSH  
       }

\address{Indian Institute of Technology Patna, Patliputra Colony \\
         Patna 800013, India \\
sghosh@iitp.ac.in}

\author{IRENE MARZOLI}

\address{School of Science and Technology, Universit\`a di Camerino \\Via Madonna delle Carceri 9,
62032 Camerino, Italy\\
irene.marzoli@unicam.it}

\maketitle

\begin{history}
\received{Day Month Year}
\revised{Day Month Year}
\end{history}

\begin{abstract}
The time evolution of a particle, caught in an infinitely deep
square well, displays unexpected features, when one includes tiny
relativistic effects. Indeed, even the smallest corrections to the
non-relativistic quadratic spectrum manifest themselves in a
dramatic way. Our theoretical analysis brings to light a
completely new time scale, at which the system exhibits
surprisingly perfect revivals. This longer time scale rules the
system dynamics and replaces the original revival time of the
unperturbed system. The early manifestation of such phenomenon
is captured by the sensitivity of sub-Planck structures
for different values of the relativistic
corrections.
\end{abstract}


\section{Introduction}

The dynamics of wave packets features a wealth of interesting
effects known as collapses and revivals, both integer and
fractional.\cite{averbukh,robinett1,berry}
These phenomena have
been investigated theoretically and observed experimentally in a
variety of physical systems, including Rydberg wave packets,\cite{parker,alber,ahn}
molecular systems,\cite{zewail,stolow,garraway,katsuki,ohmori,katsuki2} wave
packets in semiconductor quantum wells,\cite{leo,koch} photon
cavity systems,\cite{rempe} and Bose-Einstein condensates.\cite{wright,choi,ruostekoski}
Even an optical analogue of a
quantum particle, bouncing on a hard surface under the influence
of gravity, a so-called quantum bouncer, has been experimentally
realized,\cite{dellavalle} thus allowing for the observation
of fractional and integer revivals.

Most of the theoretical analysis, available in the literature, is
restricted to systems with a spectrum, which depends quadratically
on the quantum number $n$. In this case, the revival dynamics
becomes exact. The prototypical example is a particle in an
infinitely deep square well. This system has been studied in great
detail in the context of the so-called \emph{quantum carpets},
i.e. highly regular space-time structures in the probability
density,\cite{marzoli1,kaplan1} as well as in phase space by
means of the Wigner distribution.\cite{stifter,friesch,robinett2}

There are few examples, however, where quantum revivals have been investigated
in the relativistic regime. In this context, an analysis has been
made for a slightly relativistic particle caught in an infinitely
deep square well.\cite{marzoli}
This treatment takes into account
the relativistic effects as perturbations, that produce a quartic
correction to the non-relativistic energy spectrum. For very small
values of the relativistic parameter, the relativistic corrections
determine a shift in the characteristic revival time of the
system, whereas larger values completely wash out the highly
regular spatio-temporal patterns in the probability density.
Recently, attention has been paid to the fully
relativistic case, by solving Dirac equation for a particle
confined to a circumference of radius $R$.\cite{strange} This
analysis has confirmed that generally revivals do not occur within
a relativistic regime. Only carefully designed wavepackets, with
specific mathematical properties, can display quantum revivals and
weave \emph{quantum carpets} within a fully relativistic theory.

Here, we investigate more carefully the role and the effects of
the relativistic corrections for a slightly relativistic particle
in a box. We find out that the modifications to the quadratic
spectrum of the system result in a new revival time scale, giving
rise to super revivals.\cite{bluhm,anu} In particular, we focus on
the dependence of the energy eigenvalues on the fourth power of
the quantum number. This feature suggests the possible existence
of a different and new time scale, governing the system dynamics.
Surprisingly, even in the slightly relativistic regime, this
modest modification can produce large nonlinear effects on a
sufficiently long time scale. Our theoretical analysis clearly
demonstrates that a perfect revival of the original wave packet
takes place at a definite time $T_{\mathrm{sr}}^{(4)}$, depending
on the relativistic effects. This revival manifests itself both in
the space-time representation of the probability density and in
phase space. Here, we can appreciate even the smallest features of
the Wigner function, the so-called sub-Planck scale
structures.\cite{zurek,pathak,toscano} These highly nonclassical
features are extremely sensitive to perturbations and decoherence,
as pointed out in a variety of
systems.\cite{ghosh1,praxmeyer,jay,bhatt,milburn,scott,ghosh2,roy}
Therefore, they seem to be an ideal tool to monitor and
characterize the Wigner function at or nearby a fractional revival
time. To provide quantitative evidence to our predictions, we
calculate the relevant sub-Planck dimension for the Wigner
function at typical fractional revival times. The sensitivity of
these interference structures allows us to compare even the finest
details of the wave packet dynamics for different values of the
relativistic corrections. This quantitative measure provides a
clear proof that the system evolution exhibits full and fractional
revivals on a new time scale $T_{\mathrm{sr}}^{(4)}$, set by the
relativistic effects. These revivals reproduce all the features of
the original wave packet, exactly as it happens in the unperturbed
non-relativistic case.

The paper is organized as follows. In Sec.~\ref{box}, we briefly
review the basics of a slightly relativistic particle in a box.
The rich structure of quantum carpets and the phase space Wigner
distribution are presented and discussed in
Sec.~\ref{revival-dynamics}, where we analyse the wave packet
dynamics. Both approaches clearly show the existence of a new
revival time for the system, due to the relativistic corrections.
To quantify differences and similarities between the Wigner
functions for different values of the relativistic corrections, we
consider the variation of sub-Planck dimension for these phase
space structures. A sensitivity analysis of these structures is
discussed in Sec.~\ref{subPlanck}. Finally, we conclude by
summarizing our main results in Sec.~\ref{conclusions}.
\section{\label{box}
         Slightly relativistic particle in an infinite square well}
In ordinary laboratory conditions, the electronic energy is
of the order of few eV. Hence, one would expect that corrections, much smaller than
this order of magnitude, should be negligible.
However,
such a little modification can produce a large nonlinearity and
introduce a new time scale in the dynamics of a slightly
relativistic particle.

The relativistic Hamiltonian of a particle of mass $m$ is given by
\begin{equation} \label{rel_ham}
H = \sqrt{m^{2}c^{4}+p^{2}c^{2}}-mc^{2}.
\end{equation}
To our purposes it suffices to consider an approximate version of
this Hamiltonian, by making an expansion of Eq.~(\ref{rel_ham})
\begin{equation} \label{ham}
  H \simeq \frac{p^2}{2m} - \frac{1}{2mc^2}
                            \left(\frac{p^2}{2m}
                            \right)^2.
\end{equation}
We note that Hamiltonian, Eq.~(\ref{ham}), includes quartic
corrections to the non-relativistic Hamiltonian of a particle in
an infinitely deep square well. The corresponding eigenvectors and
eigenvalues satisfy the following time independent Schr\"{o}dinger
equation for a slightly relativistic particle
\begin{equation} \label{schroe}
  H |u_{n}\rangle= E_{n} |u_{n}\rangle.
\end{equation}
Enforcing the boundary conditions at the walls of the box located
at $x=0$ and $x=L$
\begin{equation}
u_{n}(x=0)=u_{n}(x=L)=0
\end{equation}
and trying the following \emph{ansatz}
\begin{equation}
u_{n}(x) = \sin(k_{n}x) \,,
\end{equation}
the Schr\"{o}dinger equation (\ref{schroe}) reads
\begin{equation}
  \left[\frac{\hbar^2k_{n}^2}{2m}-\frac{1}{2mc^2}\left(\frac{\hbar^2k_{n}^2}{2m}
  \right)^2-E_{n}\right] \sin(k_{n}x)=0 \, ,
\end{equation}
where $k_{n}=n\pi/L$. It leads to the dispersion relation
\begin{equation} \label{energy-levels}
E_{n}=\frac{\hbar^2k_{n}^2}{2m}-\frac{1}{2mc^2}
      \left(\frac{\hbar^2k_{n}^2}{2m}\right)^2 ,
\end{equation}
with the normalized energy eigen functions of the slightly
relativistic particle being
\begin{equation}
u_{n}(x)=\sqrt{\frac{2}{L}} \sin\left(n\pi\frac{x}{L}\right).
\end{equation}
In the non-relativistic case, the spectrum is quadratic in the
quantum number $n$ and the energy levels are all integer multiples
of the lowest eigenvalue. Therefore, the system dynamics is
periodic with a characteristic revival time defined as
\begin{equation} \label{T1}
T_{\mathrm{rev}} \equiv \frac{4m L^2}{\pi\hbar}.
\end{equation}
If we insert the revival time, Eq.~(\ref{T1}), in the expression,
Eq.~(\ref{energy-levels}), for the energy eigenvalues, we obtain
\begin{equation} \label{energy}
E_{n}= \left( n^2-q^2n^4 \right) \, \hbar\frac{2\pi}{T_{\mathrm{rev}}} \,,
\end{equation}
where we introduced the relativistic parameter
\begin{equation}
q\equiv \frac{1}{4L} \left(\frac{2\pi\hbar}{mc}\right)
 \equiv\frac{\lambda_{c}}{4L} \,,
\end{equation}
as the ratio between the Compton wavelength $\lambda_{c}$ of the
particle and the box width $L$. The parameter $q$ controls the
size of the relativistic effects. For an electron, the Compton
wavelength is approximately equal to $2.43 \times 10^{-12}$~m.
Therefore, even in nanostructures, like quantum dots, realistic
values of the parameter $q$ are of the order of $10^{-3}$.

Despite the smallness of the relativistic parameter $q$, the
quartic term in the expression, Eq.~(\ref{energy}), for the energy
eigenvalues leads to a new revival time, given by
\begin{equation} \label{T2}
T_{\mathrm{sr}}^{(4)}=\frac{4 m L^{2}}{q^{2} \pi \hbar}
     = \frac{T_{\mathrm{rev}}}{q^2}.
\end{equation}
This second revival time is well separated from the
non-relativistic one. Indeed, the ratio between the two revival
times $T_{\mathrm{rev}}/T_{\mathrm{sr}}^{(4)}$ is equal to the
relativistic parameter $q^2$, which is typically quite small.
Hence, the two time scales are utterly different.
\begin{figure}[htpb]
\centering
\includegraphics[width=3.2 in]{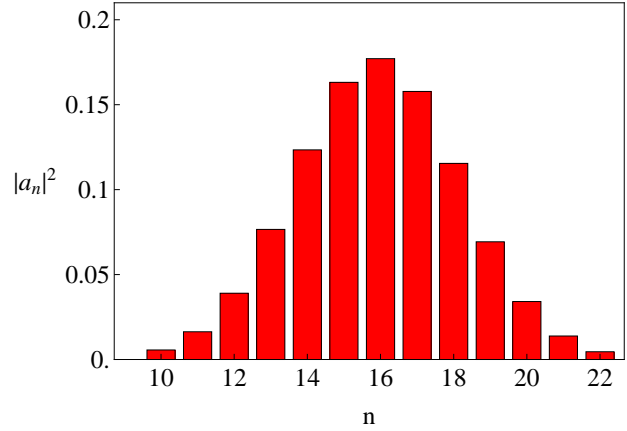}
\caption{\label{distribution} (Color online) Population
distribution $|a_{n}|^2$ as a function of the quantum number $n$
for an initial wave packet with width $\Delta x = L /10$ and
average momentum $\bar{p}=50 \hbar/L$. The distribution is peaked
around $\bar{n}=16$ with a spreading $\Delta n \simeq 6$.}
\end{figure}
\section{\label{revival-dynamics}
         Revival dynamics of a slightly relativistic particle}
\subsection{Gaussian wave packet and quantum carpets}
A particle in an infinitely deep square box can be well
approximated by a Gaussian wave packet
\begin{equation} \label{gauss}
\psi(x) =
         \frac{1}{(\sqrt{\pi}\Delta x)^{1/2}}
         \exp\left[-\frac{1}{2}\left(\frac{x-\bar{x}}{\Delta
         x}\right)^2 + i \, \frac{\bar{p} \, (x-\bar{x})}{\hbar}
             \right] ,
\end{equation}
where $\bar{x}$ and $\bar{p}$ represent, respectively, the initial
position and the average momentum. To study the time evolution of
the particle, it is convenient to expand the wave packet in terms
of the energy eigen states $|u_n\rangle$
\begin{equation} \label{probability_amp}
\psi(x,t) = \sum_{n=1}^{\infty} a_{n} u_{n}(x)
            e^{-iE_{n}t/\hbar} ,
\end{equation}
where the expansion coefficients are given by
\begin{equation} \label{an}
a_n =\frac{1}{2 i}\sqrt{\frac{4 \Delta x
\pi}{L\sqrt{\pi}}}\left[e^{i n \pi \bar{x}/L}e^{-\Delta
x^2(\bar{p}+n\pi \hbar/L)^2/2\hbar^2}-e^{-i n \pi
\bar{x}/L}e^{-\Delta x^2(\bar{p}-n\pi \hbar/L)^2/2\hbar^2}\right]
\end{equation}
and satisfy the normalization condition
\begin{equation}
\sum_{n=1}^{\infty} |a_{n}|^2 = \int_{0}^{L}|\psi(x,t)|^{2}dx = 1
.
\end{equation}
From the expression, Eq.~(\ref{an}), for the expansion
coefficients $a_n$, we see that the wave packet, describing the
particle in the box, is peaked around the energy eigenstate with
average quantum number $\bar{n} = \bar{p} L/(\hbar\pi)$. Other
energy eigenstates with a non-negligible overlap with the initial
wave packet are characterized by quantum numbers in the range
$\bar{n} \pm \Delta n$. As an example, we plot in
Fig.~\ref{distribution} the population distribution $|a_n|^2$ for
an initial wave packet with $\bar{p} = 50 \hbar/L$, that is peaked
around the quantum number $\bar{n} = 16$. In this case, the
spreading $\Delta n \simeq 6$, since it is enough to include the
energy eigenstates ranging from $n=10$ to 22 to satisfy the
normalization condition up to more than 99\%.

Equation (\ref{probability_amp}) is the starting point to
numerically investigate the probability density to find the
particle in the box at a given time and position.
\begin{figure}[htpb]
\centering
\includegraphics[width=5.6 in]{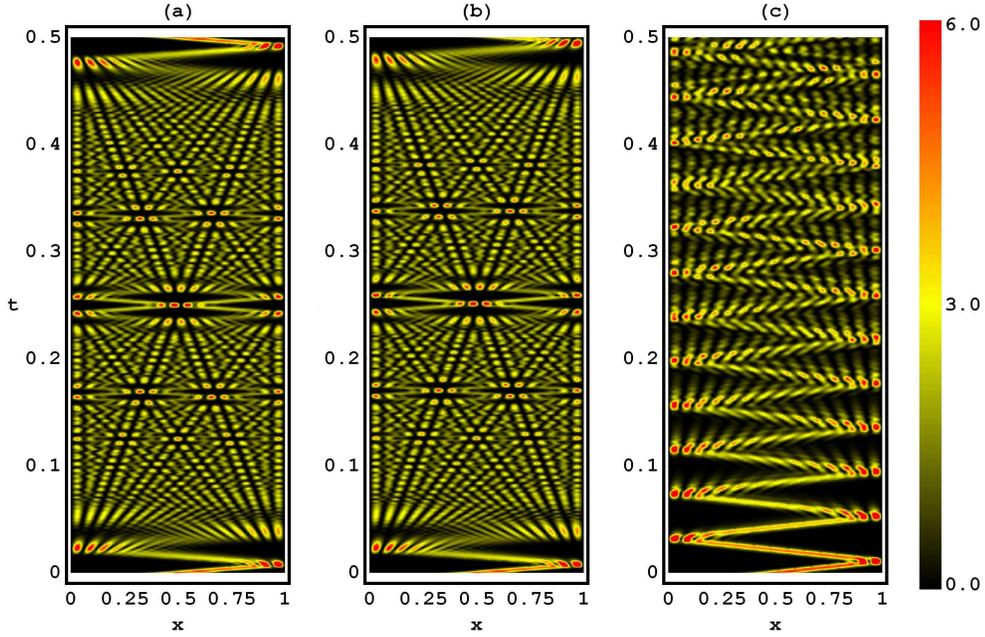}
\caption{(Color online) Plot of the probability density to find
the particle at time $t$ and position $x$ in an infinitely deep
square well. The initial Gaussian wave packet is centered at
$\bar{x}=L/2$ and has a width $\Delta x=L/10$. The average
momentum is $\bar{p}= 50 \hbar/L$ for all cases. (a) It shows the
space-time structures of a non-relativistic particle, whereas (b)
and (c) include small relativistic corrections with
$q^{2}=10^{-5}$ and $q^{2}= 5\times 10^{-4}$, respectively.
Position is expressed in units of the box length $L$, while time
is scaled by the revival time $T_{\mathrm{rev}}$.} \label{carpets}
\end{figure}
Density plots of the probability density of the evolved wave
packet are shown in Fig.~\ref{carpets}. Here, we have chosen the
wave packet initially centered at $\bar{x}=L/2$ and with average
momentum $\bar{p}= 50 \hbar / L$. The three panels present
different cases, ranging from the non-relativistic one $(q=0)$, to
others with small relativistic corrections ($q^2 = 10^{-5}$ and $5
\times 10^{-4}$). During the time evolution, the wave packet
breaks into small replicas of itself, thus giving rise to the
phenomenon of fractional revivals. Finally, it regains its
original shape at the time $T_{\mathrm{rev}}/2$. For this reason,
here we consider the time evolution only up to time
$T_{\mathrm{rev}}/2$. Fractional revivals take place at times
$t=rT_{\mathrm{rev}}/s$, with $r$ and $s$ mutually prime integers.
The scenario of full and fractional revivals characterizes case
(a), which depicts a non-relativistic particle. Also in the
presence of very small relativistic corrections, plot (b), the
main features are preserved. However, already for a slightly
larger value of the relativistic parameter $q^2 = 5 \times
10^{-4}$ (c), the overall \emph{quantum carpet} disappears, being
replaced by a more classical trajectory, which reminds of a ball
bouncing between the two walls. The relativistic corrections
completely change the space-time structures in the probability
density and reveal a more classical nature marked by periodic
oscillations. This classical behaviour, which governs the system
dynamics at earlier times, is characterized by the oscillation
period
\begin{equation} \label{tcl}
T_\mathrm{cl}=\frac{2L}{v_\mathrm{cl}}=\frac{T_{\mathrm{rev}}}{2\bar{n}}
,
\end{equation}
where the average velocity of the particle is
$v_\mathrm{cl}=\bar{p}/m$. Thus, this classical time scale
$T_\mathrm{cl}$ depends on the average value of the distribution,
i.e. $\bar{n}$. In our analysis, we have chosen $\bar{p}= 50 \hbar
/ L$ and the corresponding average quantum number is $\bar{n}=16$,
as shown in Fig.~\ref{distribution}. Interestingly, for
relativistic corrections of the order $q^2 = 5 \times 10^{-4}$,
one can observe almost $\bar{n}$ oscillations of the particle
trajectory up to the time $T_{\mathrm{rev}}/2$ [see
Fig.~\ref{carpets}(c)].

Alternatively, whenever we deal with a wave packet, characterized
by a well defined average quantum number $\bar{n}$ and
encompassing a few other energy eigen states, we can resort to a
Taylor expansion of the energy eigenvalues around $\bar{n}$
\begin{equation} \label{taylor}
E_{n}\simeq
E_{\bar{n}}+E^{'}_{\bar{n}}(n-\bar{n})+\frac{1}{2!}E^{''}_{\bar{n}}(n-\bar{n})^2
       +\frac{1}{3!}E^{'''}_{\bar{n}}(n-\bar{n})^3+ \ldots,
\end{equation}
where primes denote derivatives with respect to $n$.
The coefficients of the Taylor series are then directly related to
the relevant time scales of the wave packet evolution. This
method, when applied to the energy eigen values,
Eq.~(\ref{energy}), for a slightly relativistic particle, not only
amends the time periods characterizing the dynamics of the
non-relativistic case, but also introduces new time scales. In
particular, the classical time scale $T_\mathrm{cl}$, which
controls the initial behavior of the wave packet, becomes
\begin{equation} \label{tcl_bis}
\bar{T}_\mathrm{cl}=\frac{2\pi \hbar}{|E^{'}_{\bar{n}}|}
                   =\frac{T_{\mathrm{rev}}}{2\bar{n} - 4 q^2\bar{n}^3}
                   = \frac{T_\mathrm{cl}}{1 - 2 q^2 \bar{n}^2}.
\end{equation}
This expression reveals that $\bar{T}_\mathrm{cl}$ is slightly
larger than $T_\mathrm{cl}$, as $q^2$ is a very small quantity.
Similarly, the quadratic term in Taylor series expansion yields a
new revival time
\begin{equation} \label{tclnew}
\bar{T}_\mathrm{rev}=\frac{2\pi
\hbar}{\frac{1}{2}|E^{''}_{\bar{n}}|}=\frac{T_{\mathrm{rev}}}{1-6
q^2\bar{n}^2}.
\end{equation}
Again the relativistic corrections lead to an increase of the
revival time in comparison to the non-relativistic case. This
small shift is visible in Fig.~\ref{carpets}(b), where the wave
packet is not yet back at the initial position $\bar{x} = L/2$ at
exactly $T_{\mathrm{rev}}/2$. A small extra time is required to
observe the full and complete revival.

Moreover, super revival times appear in correspondence with the
cubic and quartic terms of Taylor series
\begin{equation} \label{tsuper1}
     T_{\mathrm{sr}}^{(3)}=\frac{2\pi \hbar}{\frac{1}{6}|E^{'''}_{\bar{n}}|}
                     =\frac{T_{\mathrm{rev}}}{4 \bar{n} q^2}
\;\;\;\mathrm{and}\;\;\;
     T_{\mathrm{sr}}^{(4)}=\frac{2\pi \hbar}{\frac{1}{24}|E^{''''}_{\bar{n}}|}
                                =\frac{T_{\mathrm{rev}}}{q^2} .
\end{equation}
These two time scales $T_{\mathrm{sr}}^{(3)}$ and
$T_{\mathrm{sr}}^{(4)}$ both depend on the relativistic parameter
$q^2$. When $q^2$ is very small, these two times are much greater
than the revival time $\bar{T}_{\mathrm{rev}}$. In this case, the system
time evolution is still controlled by $\bar{T}_{\mathrm{rev}}$, while
$T_{\mathrm{sr}}^{(3)}$ and $T_{\mathrm{sr}}^{(4)}$ are not
yet playing a significant role. On the contrary, for a
non-negligible relativistic parameter as in Fig.~\ref{carpets}(c),
a drastic change takes place in the system behavior. Classical
oscillations persist for a comparatively longer time, at least of
the order of $T_{\mathrm{rev}}/2$. Under these conditions,
$T_{\mathrm{sr}}^{(3)}$ and $T_{\mathrm{sr}}^{(4)}$ rule the
system dynamics and govern the appearance of the usual fractional
and full revivals. Hence, the existence of the super revival time
scales appears only for relativistic corrections. We will
investigate further this behaviour in our phase space analysis.
\subsection{Phase space Wigner distribution}
\begin{figure}[htpb]
\centering
\includegraphics[width=\textwidth]{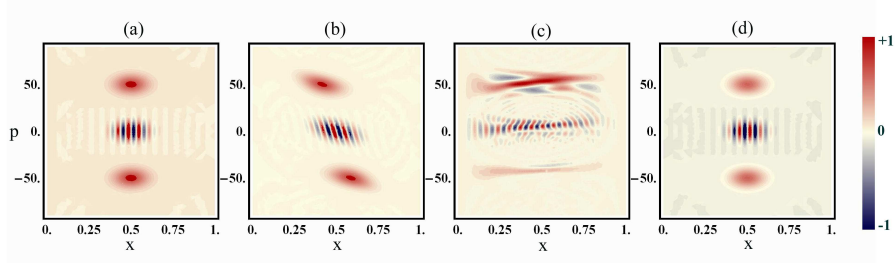}
\caption{(Color online) Snapshots of the Wigner distribution
describing a particle in an infinitely deep square well. The
initial wave function is a Gaussian wave packet centered in
$\bar{x} = L/2$ with average momentum $\bar{p} = 50 \hbar/L$. (a)
Non-relativistic particle at time $t=T_{\mathrm{rev}}/4$; (b) same
time but for small relativistic corrections with $q^2=10^{-5}$ and
(c) comparatively larger relativistic corrections with $q^2=
5\times 10^{-4}$. A slightly relativistic particle ($q^{2}= 5
\times 10^{-4}$) shows again a perfect two-way break up but on a
longer time scale at time $t=T_{\mathrm{sr}}^{(4)}/4$, as found in
(d).} \label{wigner}
\end{figure}

An alternative tool to investigate full and fractional revivals is
represented by the Wigner function, which provides a complete
description of these phenomena in phase space. Moreover, the
negativity of the Wigner function is a hallmark of the
non-classicality of quantum states. The Wigner function is defined
as
\begin{equation}
W(x,p,t) = \frac{1}{\pi \hbar} \int_{-\infty}^{+\infty} \psi^{*}(x
-x',t) \, \psi(x+x',t) \, e^{-2ipx'/\hbar} dx'.
\end{equation}
With our choice of an initial Gaussian wave packet,
Eq.~(\ref{gauss}), the corresponding Wigner function at time $t=0$
is
\begin{equation}
W(x,p)=\frac{1}{\pi\hbar} \frac{1}{\sqrt{2\pi}}
\exp\left[-\left(\frac{x-\bar{x}}{\Delta x}\right)^2\right]
\exp\left[-\left(\frac{p-\bar{p}}{\hbar / \Delta
x}\right)^2\right].
\end{equation}
It is a Gaussian distribution, centered at $\bar{x}$ and $\bar{p}$
in phase space. During its time evolution, the wave packet
undergoes a spreading, before reviving at characteristic times.
For instance, the well known Schr\"{o}dinger cat state will appear
at $t=T_{\mathrm{rev}}/4$, when the wave packet consists of two
components with opposite momenta. This feature is apparent in the
corresponding Wigner function, as depicted in
Fig.~\ref{wigner}(a). The wave function is the superposition of
two Gaussian wave packets with the same average position but
different average momenta $\bar{p}$ and $-\bar{p}$, respectively.
The density plot of the Wigner function clearly shows the two-way
breakup in phase space and the interference ripples appearing in
the middle. This scenario is modified when dealing with a slightly
relativistic particle. A small relativistic correction
$q^{2}=10^{-5}$, same as that chosen in Fig.~\ref{carpets}(b), is
considered in the case of Fig.~\ref{wigner}(b). It depicts a small
deviation from the expected fractional revival, which is not
clearly visible in the carpet structure of Fig.~\ref{carpets}(b).
Figure~\ref{wigner}(c) shows the density plot of the Wigner function
again at time $t=T_{\mathrm{rev}}/4$ for a comparatively larger
relativistic correction, $q^2= 5 \times 10^{-4}$, same as that
chosen for the probability density plot of Fig.~\ref{carpets}(c).
This small correction completely destroys the two-way breakup
structure.

The reason, behind this phenomenon, is that the revival time,
$T_{\mathrm{rev}}$, comes from the quadratic term in the
expression, see Eq.~(\ref{energy}), for the energy eigenvalues.
However, the relativistic corrections can produce large
non-linearities of higher order, which introduce new time scales,
Eq.~(\ref{tsuper1}).
So the wave packet motion for a slightly relativistic particle is
governed by the super revival times $T_{\mathrm{sr}}^{(3)}$
and $T_{\mathrm{sr}}^{(4)}$. The system will revive at times,
which are multiple of both $T_{\mathrm{sr}}^{(3)}$ and
$T_{\mathrm{sr}}^{(4)}$.

Therefore, a fractional revival
will take place at time $t_{\mathrm{frac}}$ provided that
\begin{equation} \label{fractional}
t_{\mathrm{frac}}  %
                 =\frac{r_1}{s_1} T_{\mathrm{sr}}^{(3)}
                 =\frac{r_{2}}{s_{2}}T_{\mathrm{sr}}^{(4)},
\end{equation}
where $r_{1}$, $s_{1}$, and $r_{2}$, $s_{2}$ are relative prime
integers. For very small relativistic corrections, such as
$q^2=10^{-5}$, just a deviation from the non-relativistic
behaviour is observed on the shorter time scale, at times $t$ of
the order of the first revival time $T_{\mathrm{rev}}$. So, in
this range of relativistic corrections, the first revival time is
still playing a role in the overall system dynamics. Moreover, the
super revival time $T_{\mathrm{sr}}^{(4)}= 4 \bar{n}
T_{\mathrm{sr}}^{(3)}$ defines the full and perfect revival
of the system state. For instance, our numerical investigations
predict a perfect two-way break up of the Wigner function at time
$T_{\mathrm{sr}}^{(4)}/4$, also when the relativistic parameter is
just $q^2 = 10^{-5}$. If we further increase the relativistic
parameter $q^2$, as shown in Fig.~\ref{wigner}(c), on the short
time scale, around $T_{\mathrm{rev}}$, the system dynamics is
completely different from the non-relativistic case. Apparently
the ordered space-time structures of quantum carpets are lost [see
Fig.~\ref{carpets}(c)]. To recover the periodic time behaviour we
should wait for a longer time. In this case, the system dynamics
is dominated by the super revival time. Fractional and full
revivals take place only on the longer time scale, set by
$T_{\mathrm{sr}}^{(4)}$, as demonstrated in Fig~\ref{wigner}(d).

The explanation behind this effect is the competition between the
two time scales. For non-negligible relativistic corrections, the
relevant time scale is set by $T_{\mathrm{sr}}^{(4)}$. From
Eq.~(\ref{T2}) it is evident that whenever the system satisfies
the revival condition for $T_{\mathrm{sr}}^{(4)}$, automatically
also the other condition, involving $T_{\mathrm{rev}}$ is
satisfied. The vice versa is not true and fractional revivals on
the shorter time scale disappears. For a non-relativistic particle
the unique revival time is $T_{\mathrm{rev}}$. At intermediate
regimes, there is a coexistence of both the revival times
$T_{\mathrm{rev}}$ and $T_{\mathrm{sr}}^{(4)}$, whereas for
slightly larger relativistic effects the dynamics is dominated by
$T_{\mathrm{sr}}^{(4)}$ only.

\begin{figure}[htpb]
\centering
\includegraphics[width=2.6in]{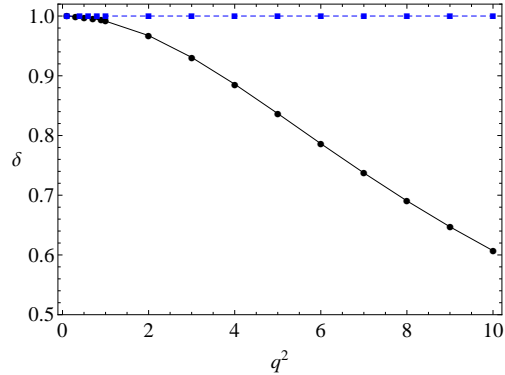}
\caption{(Color online) Sensitivity of sub-Planck structures, in
phase space, as a function of the relativistic parameter $q^2$,
here in unit of $10^{-6}$. In the vertical axis we report the
ratio $\delta \equiv a_q /a$ between the sub-Planck dimension in
the relativistic case and the non-relativistic one ($q^2 = 0$)
calculated at $t=T_{\mathrm{rev}}/4$. The solid line joins the
full dots calculated at $T_{\mathrm{rev}}/4$, whereas the dashed
line joins the points pertaining to the case at
$T_{\mathrm{sr}}^{(4)}/ 4$.} \label{sensitivity}
\end{figure}

\section{\label{subPlanck}
         Sensitivity of sub-Planck structures in phase space}
Phase space interference structures of dimension smaller than
$\hbar$, called sub-Planck scale structures, are well studied in
the context of their sensitivity to decoherence and pertubations.
Their dimension is defined as $a \approx \hbar^2/A$, where $A$
represents the classical action. Interestingly, these structures
give an indication of the system sensitivity against
perturbations. For this reason they appear to be an appropriate
tool to quantify the modifications, due to the relativistic
effects, of phase space structures in the Wigner function of a
particle in the box. For instance, they can provide a quantitative
measure of the differences between the Wigner functions of
Figs.~\ref{wigner}(a) and (b), pertaining, respectively, to the
non-relativistic case ($q^2=0$) and to a slightly relativistic one
($q^2=10^{-5}$).

To this end, we focus on the fractional revival appearing at time
$T_{\mathrm{rev}}/4$ and characterize the interference fringes
through their sub-Planck dimension. Our numerical analysis yields
the classical action for the non-relativistic particle being
$A\approx\Delta x\Delta p=3.57 \hbar$ and the corresponding
sub-Planck structure size is $a=0.28 \hbar$. As shown in
Fig.~\ref{wigner}(b), at the time $T_{\mathrm{rev}}/4$, the
two-way break up of the Wigner function is also present when the
relativistic parameter is just $q^2 = 10^{-5}$. This small
relativistic correction causes a perturbation in the system, which
results in a change of sub-Planck dimension $a_{q}=0.17 \hbar$.
Figure~\ref{sensitivity} shows the variation of sub-Planck
dimension, at $t= T_{\mathrm{rev}}/ 4$ (solid line), as a function
of the relativistic parameter $q^2$. In particular, we introduce
the quantity $\delta \equiv a_q / a$, which provides the ratio of
sub-Planck dimension of the slightly relativistic case to the
non-relativistic one. The plot takes into account values of the
relativistic parameter up to $q^2 = 10^{-5}$. It clearly depicts
the fall of sub-Planck dimension of the interference fringes with
the increase of relativistic effects. Thus, even the smallest
relativistic corrections are captured by the size of sub-Planck
structures.

It is then interesting to compare the fractional revival [see
Fig.~\ref{wigner}(d)], taking place at $T_{\mathrm{sr}}^{(4)}/4$
in the slightly relativistic case, to the one at
$T_{\mathrm{rev}}/4$ for the non-relativistic particle [see
Fig.~\ref{wigner}(a)]. Hence, we have calculated the ratio between
the sub-Planck dimensions of the interference fringes of the
Wigner function at these two fractional revival times. The results
are presented in Fig.~\ref{sensitivity} (dashed line): the
sub-Planck dimension remains constant for all the values of the
relativistic parameter $q^2$. This is a quantitative evidence
that, in the slightly relativistic case, the wave packet dynamics
not only revives on the time scale set by $T_{\mathrm{sr}}^{(4)}$,
but maintains all the characteristics of fractional and full
revivals typical of the non-relativistic case.
\section{\label{conclusions} Conclusions}
Relativistic corrections to the energy spectrum of a particle in
an infinitely deep square box not only amend the typical revival
time but, most notably, introduce a super revival on a much longer
time scale. Hence, quartic terms in the energy spectrum of a
slightly relativistic particle are not simply a higher-order
correction, but play a definite role in the system dynamics. For
very small values of the characteristic relativistic parameter,
the system time evolution still exhibits full and fractional
revivals at the expected times. However, one can perceive some
distortions both in the regular pattern of space-time probability
density, the so-called \emph{quantum carpets}, and in the phase
space representation of Wigner function. These deviations are better
captured by the sensitivity of sub-Planck structures.

For slightly larger values of the relativistic parameter, there is
a dramatic change in the overall picture. The particle probability
density does not spread but gives rise to a more or less localized
trajectory, reminiscent of a classical ball bouncing between the
two walls. Interference effects seem to disappear even at moderate
relativistic regimes. This behaviour obscures the expected full
and fractional revivals on a time scale of the order
$T_{\mathrm{rev}}$. However, the system dynamics is now governed
by a new time scale, $T_{\mathrm{sr}}^{(4)}$, which depends on the
quartic terms in the energy spectrum and is only due to
relativistic effects. The familiar full and fractional revivals,
that mark the time evolution of the non-relativistic particle,
manifest themselves on this new time scale
$T_{\mathrm{sr}}^{(4)}$. Again the analysis of sub-Planck
dimension of the relevant interference fringes in the Wigner
function allows for a comparison between the fractional revivals
for different values of the relativistic parameter. The sub-Planck
scale structures provide a quantitative evidence for the perfect
reshaping of the original wave packet around
$T_{\mathrm{sr}}^{(4)}$.

In conclusion, the slightly relativistic particle in a box
exhibits super revivals at a time $T_{\mathrm{sr}}^{(4)}$, which
is set by the relativistic effects and has no counterpart in the
non-relativistic case. Remarkably, the observation of these super
revivals does not require a specially tailored or designed wave
packet, as it happens, instead, in the fully relativistic case
considered in Ref.~\refcite{strange}.

These effects may be observed in a cyclotron maser or gyrotron
based on slightly relativistic mass effect of the free electron in
vacuum.\cite{schneider} Also in narrow-gap semiconductors,
conduction electrons exhibit a pseudo-relativistic behavior.\cite{kane}
Indeed, the dispersion relation between the
conduction-band energy and the momentum closely resembles the
expression for the energy of a relativistic particle, in which the
speed of light $c$ is replaced by a characteristic velocity,
determined by the energy gap and the effective mass of the
conduction electron. Therefore, our considerations
apply to these systems
as well.

\end{document}